\documentclass[12pt]{revtex4}
\usepackage{epsfig}

\newcommand{\avrg}[1]{\langle{#1}\rangle}
\newcommand{\ql}{\ell}
\newcommand{\qlc}{\ell_{\rm c}}
\newcommand{\width}{\Delta\ell}
\newcommand{\muS}{\mu_{\rm s}}
\newcommand{\muD}{\mu_{\rm d}}
\newcommand{\eg}{{\em e.g.}}
\newcommand{\yieldstress}{\sigma_{\rm c}}

\begin{document}

\title{Static versus dynamic friction: The role of coherence}
\author{Z\'en\'o Farkas\dag, S\'{\i}lvio R. Dahmen\ddag \, and D. E. Wolf\dag}
\affiliation{\dag\ Department of Physics, Universit\"at
Duisburg-Essen, D-47048 Duisburg, Germany}
\affiliation{\ddag\ Instituto de F\'{\i}sica da UFRGS, CP 15051,
90501--970 Porto Alegre RS, Brazil}

\begin{abstract}
A simple model for solid friction is analyzed. It is based on
tangential springs representing interlocked asperities of the surfaces
in contact. Each spring is given a maximal strain according to a
probability distribution. At their maximal strain
the springs break irreversibly. Initially all springs are assumed to
have zero strain, because at static contact local elastic stresses are
expected to relax. Relative tangential motion of the two solids leads
to a loss of coherence of the initial state: The springs get out of
phase due to differences in their sizes. This mechanism alone is shown to
lead to a difference between static and dynamic friction forces already. We
find that in this case the ratio of the static and dynamic
coefficients decreases
with increasing relative width of the probability distribution, and
has a lower bound of $1$ and an upper bound of $2$.
\end{abstract}

\pacs{81.40.Pq, 46.55.+d}
\maketitle

While the facts that dry solid friction is proportional to the normal
load at the contact and does not depend on the apparent contact area
were established experimentally at least as early as in the 16th
century by Leonardo da Vinci and are now known under the names of
Amonton (1699) or Coulomb (1781) \cite{Bowden1958}, it was probably
Euler (1750) who first distinguished between static and dynamic
friction \cite{Duran2000}.  This difference has been explained in
several, conceptually different ways. The reason was identified as: A
collective depinning phenomenon \cite{Volmer1997}, the time
strengthening of individual pinning sites
\cite{Caroli1997a,Caroli1997}, the shear
melting of a lubrication film \cite{Persson1996}, mobile impurities at
the interface \cite{Mueser2001}, or the formation and healing of
microcracks \cite{Gerde2001}.
The fact that all these mechanisms lead to the same macroscopic
phenomenology raises the question whether they can be classified in
terms of more abstract concepts.

An attempt in this direction was made by
Caroli and Nozi\`eres \cite{Caroli1998}, who
proposed a model for dry solid friction
based on the following physical picture:
The surfaces have randomly distributed asperities which get
interlocked.  These interlocked asperities act as
pinning sites resisting tangential motion. Under tangential load they
are deformed up to a threshold (which these authors call ``spinodal
limit''), where they break
irreversibly releasing their energy in the form of phonons
into the bulk.  The threshold is a measure for the pinning
strength. They argue that their model does not lead to a difference
between static and dynamic friction, unless the strain of the pinning
sites has different statistics in the static and in the sliding case
or aging is taken into account. The latter aspect has been further
investigated in ref.~\cite{Caroli1997} and explains also the
experimentally observed time strengthening and velocity weakening of
the pinning sites.

\begin{figure}
\centerline{\epsfig{figure=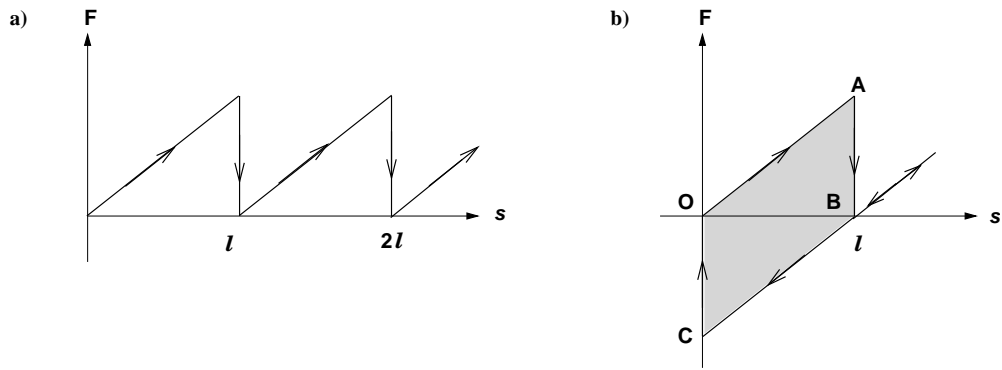,width=0.8\linewidth}}
\caption{\textbf{a)} Schematic plot of the force $F$ as a function of displacement $s$ is presented. At the threshold $\ell$
the spring breaks and immediately reattaches. \textbf{b)} By reversing the displacement the force also changes sign.
During the complete cycle $O\rightarrow A\rightarrow B\rightarrow C\rightarrow O$
energy is dissipated irreversibly and is numerically equal to the shaded area. The size of the hysteresis loop is proportional to the
threshold length $\ell$.}
\label{fig:hysteresis}
\end{figure}

However, time strengthening is a slow process. This motivates us to
explore in more detail what would be the influence of strain
statistics at the pinning sites on the static and dynamic friction
coefficients, $\muS$ and $\muD$. Although time strengthening will not
be considered, it can be included in addition to account for, \eg,
velocity weakening.


The model we consider in the following captures, we believe, the
essence of the physical picture described above and at the same time
highlights the concept of coherence, which is an ingredient in several
different models (see \cite{Mueser2003} and references therein).  For
the sake of clarity and
analytical tractability we work out only a one-dimensional version,
but the extension to the two-dimensional case is straightforward.

Consider two solid bodies in contact, one being the fixed substrate
(the ``track'') and the other the body to be displaced (the
``slider''). To keep the equations simple, we consider only motion in
a fixed direction. Reversing the direction would lead to hysteretic
behaviour like in Fig.~\ref{fig:hysteresis}.  The friction force arises from
interlocked asperities in the contact area, which are modelled here by
linear springs with zero equilibrium length (see also
\cite{Brilliantov1996}), which only act in the tangential
direction. Each spring has one end attached to the slider, while the
other end is attached to the track. When the slider moves, each spring
gets stretched up to an individual threshold length $\ql$, where it
breaks. The elastic energy stored in a spring is completely
dissipated when it breaks.

In contrast to previous work \cite{Brilliantov1996} we
take here explicitly into account that the interlocked asperities are
characterized by different threshold lengths $\ql$ with a probability
distribution $p(\ql)$,
normalized as $\int_0^\infty p(\ql) \, \ d \ql = 1$.  There is
experimental evidence that this distribution is approximately Gaussian
centered around a characteristic length \cite{Greenwood1992}.

In general the number of pinning sites and the distribution of their
strength $p(\ell)$ will change with time during a transient until
steady state sliding is reached. However, one can imagine experimental
situations where both are time independent, at least in an average sense
\cite{Dahmen2005}. Here we make this assumption deliberately in order
to show that
a difference between static and dynamic friction can ensue, even if
the number of pinning sites and the distribution of their strength are
time independent.
With this assumption a new spring with the
same parameter $\ql$ has to become active whenever one breaks. Hence
the elastic restoring force from springs of threshold length $\ql$ becomes
simply a sawtooth-like function of the displacement $s$ (see
Fig.\ref{fig:hysteresis}).  This
displacement is assumed to be the same for all springs (approximation
of a rigid slider). The friction
force is the sum of all these elastic restoring forces.  In the
following all spring constants $k$ are assumed to be the same, but
this is not crucial. For example we checked that spring constants
proportional to $\ql$ give qualitatively the same results.

The crucial ingredient of our model will be discussed now. During
sliding all springs will be stretched in the sliding direction by a
random fraction of their threshold lengths. When the motion is
stopped, the slider will recoil so that some springs get stretched in
the opposite direction until the net force on the slider is zero (in
the absence of an external shear force). In contrast to
\cite{Caroli1998} we assume here that then the strain distribution
becomes narrower, because the springs have time to relax. We have to
discard plastic flow as the main relaxation
mechanism when relative motion comes to a halt, since the speed of
this process is proportional to the difference between applied stress
and yield stress and should be therefore too slow. A different
mechanism is required, which is slow compared to
the life time $\ell/v$ of stretched springs during sliding with velocity
$v$, but fast
compared to the time a stationary contact is at rest. One possibility
might be viscoelasticity: During sliding a nonequilibrium density of
point defects in the immediate neighborhood of the surface is
created.
These point defects can be viewed as a viscous fluid penetrating
the crystal lattice: if the lattice is exposed to time dependent
stresses at very high frequencies during sliding, the defects hardly
have time to diffuse and contribute to stress relaxation. However, if
the frequency is low or even zero, as in the static case, the point
defects move due to thermal activation to regions where they reduce
the elastic energy of the entangled asperities (we note that this is
different from plastic flow, which is due to the motion of
dislocations). Diffusion of point defects is also a slow process, but
since the distances are at the nm to $\mu$m scale, and since it may be assisted
by strain, which can considerably reduce activation energies, it is
still faster than plastic deformation. Consequently we expect that
microscopic interface strains relax relatively fast in the static, but
not fast enough in the sliding case. Conceptually this is different
from time strengthening, where atomic diffusion would shift the
threshold lengths $\ell$ towards larger values, an effect that occurs
in addition, but is neglected here for the sake of working out the
effect of stain coherence more clearly.

In the context of our model we actually consider the
extreme case, where all springs relax to zero elastic energy,
as soon as sliding stops (full ``coherence'').
With the assumption that all springs are relaxed initially,
the friction force as a function of displacement $s$
for an apparent (macroscopic) contact
area  $A$ and the number density $n_{\rm p}$ of active pinning sites
(or springs) reads
\begin{equation}
  f(s) = A\, n_{\rm p}\, k\int\limits_0^\infty p(\ql) \, t(\ql,s) \,
  \ d \ql,
  \label{eq:fs}
\end{equation}
where $t(\ql,s) = s \;\mathrm{mod}\; \ql$ is a sawtooth-shaped
function of periodicity $\ql$. The phase of this periodic function is
$\phi(\ql,s) \equiv t(\ql,s)/\ql$, which is a number between 0 and
1. The behaviour of eq.~(\ref{eq:fs}) is
closely related to the probability distributions of these phases,
\begin{equation}
w(\phi,s) = \int\limits_0^\infty p(\ql)\, \delta\left(\phi -
\frac{t(\ql,s)}{\ql}\right)\, \ d \ql.
\label{eq:w}
\end{equation}
For example $w(0,s)$ is the probability density of springs that break
at displacement $s$. As these springs still contribute their elastic
restoring force to $f(s - ds/2)$ but no longer to $f(s + ds/2)$, the
derivative of eq.~(\ref{eq:fs}) is given by
\begin{equation}
f'(s)= A\, n_{\rm p}\, k \, \left[ 1 - w(0,s) \right].
\label{eq:f_prime}
\end{equation}
This can be derived from eq.~(\ref{eq:fs}) using the expression
(\ref{eq:w_sum}) given below.

Whereas the initial state is {\em coherent} in the sense that
$w(\phi,0) = \delta(\phi)$, coherence gets lost for large
displacement, where all phases become equally likely:
\begin{equation}
\lim_{s\to\infty}w(\phi,s) = 1.
\end{equation}
To prove this we evaluate the integral in eq.~(\ref{eq:w})
around each of the discrete values of $\ql$ for which the argument of
the $\delta$-function vanishes and obtain
\begin{equation}
w(\phi,s) = \sum_{m = 1}^\infty p\left(\frac{s}{m+\phi}\right)
\frac{s}{(m+\phi)^2}\ .
\label{eq:w_sum}
\end{equation}
Introducing the variable $x = (m+\phi) /s$, which becomes
quasi-continuous for large $s$, this converges to the Riemann-integral
(provided $p(\ql)$ is a Riemann-integrable function)
\begin{equation}
\lim_{s\rightarrow\infty} w(\phi,s) =  \int\limits_0^\infty
  p\left(\frac{1}{x}\right) \frac{1}{x^2}
  \ d x = \int\limits_0^\infty p(y) \ d y = 1,
\end{equation}
where the variable transformation $y = 1/x$ was used, and the last
equality is just the normalization of the distribution.  This shows
that after sufficient displacement the system forgets its initially
coherent state.  An important consequence of this {\em decoherence} is
that the friction force for large displacements becomes constant.
This follows immediately from eq.~(\ref{eq:f_prime}), because $w$
tends to 1.  Then the value of the friction force in eq.~(\ref{eq:fs})
coincides with its average,
\begin{equation}
  \avrg{f} = A\, n_{\rm p}\, \frac{k}{2} \int\limits_0^\infty p(\ql)
  \, \ql \, \ d \ql
  = A\, n_{\rm p}\, \frac{k}{2} \avrg{\ql},
\label{eq:average}
\end{equation}
where $\avrg{\ql}$ is the average maximum spring length.

\begin{figure}
\centerline{\epsfig{figure=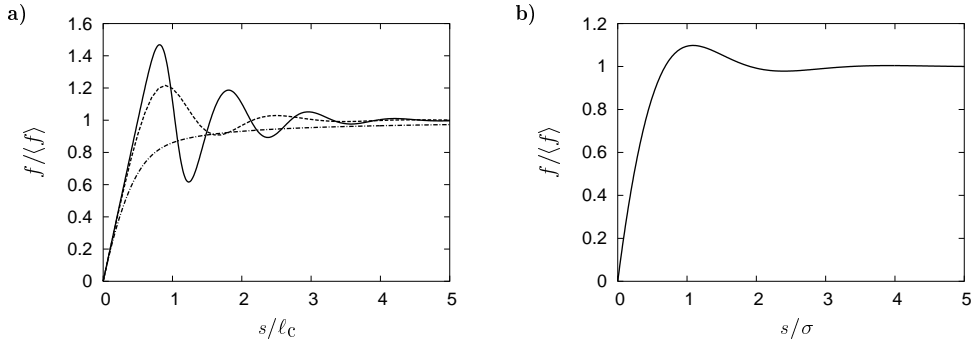,width=0.8\linewidth}}
 \caption{\textbf{a)} Friction force $f$ as a function of
  displacement $s$ of the slider for different maximum spring length
  distributions $p(\ql)$.  $\avrg{f}$ is the dynamic friction force,
  $\qlc$ is a characteristic maximum spring length. Solid line for a
  truncated Gaussian distribution, $p(\ql) \propto \exp[-(\ql-\qlc)^2
  / 2 (\width)^2]$ with width $\width = 0.15 \, \qlc$. Dashed line for
  the same Gaussian distribution, but with width $\width = 0.4 \,
  \qlc$. Dashed-dotted line for $p(\ql) \propto 1 / [1 + (\ql /
  \qlc)^3]$, width is $\width = \infty$. Note that the height of the
  first peak decreases with increasing width.  \textbf{b)} Friction
  force $f$ as a function of displacement $s$ of the slider for
  $p(\ql) \propto \exp[-\ql^2 / 2 (\width)^2]$. The maximum value of
  $f$ is approximately $1.098 \avrg{f}$ at displacement $s \approx
  1.087 \width$.}
  \label{fig:fs}
\end{figure}

Figure \ref{fig:fs}a shows the friction force $f(s)$ as a function of
displacement $s$ on a microscopic length scale, for three different
distributions $p(\ql)$. 
For the numerical evaluation it was useful to rewrite the
force $f(s)$ in eq.~(\ref{eq:fs}) by integrating eq.~(\ref{eq:f_prime}),
after formula (\ref{eq:w_sum}) has been inserted:
\begin{equation}
  f(s) = A\, n_{\rm p}\, k \, \left[s
  - \sum_{m = 1}^\infty \int\limits_0^{s/m} p(\tilde s) \, \tilde s \, 
  \ d \tilde s \right].
\end{equation}
In this expression the integrals can be calculated analytically for
simple distributions, \eg, truncated Gaussians.  All curves in
fig.~\ref{fig:fs}a have in common that they start from zero with a
slope of 2 in the natural units, $\avrg{f}$ and $\qlc$, and converge
to 1 for large displacements.  For the truncated Gaussian
distributions the force has a maximum at a displacement close to
$\qlc$, where a large number of relatively strong springs are just
about to break. As long as an external driving force remains smaller
than this maximum, the position of the slider will move only on the
scale of $\qlc$. Therefore we interpret the first
peak of the force as the {\em static} friction force, the pulling
force needed to initiate motion of the slider. Once the object moves,
the force necessary to maintain its motion decays to a smaller value
with damped oscillations, which die out again on the scale of
$\ell_{\rm c}$. Therefore we interpret the asymptotic force as the
{\em dynamic} friction force.

We found numerically that the
maximum is less pronounced the wider the Gaussian distribution
is for a given $\qlc$. This means that the difference between
static and dynamic friction decreases. An interesting question is what
happens for a distribution with finite average value, but infinite
width. An example is shown in fig.~\ref{fig:fs}a. In this case
the force monotonically increases to $\avrg{f}$, so that according to
our interpretation the static and dynamic friction coefficients are equal.
However, we regard this example purely as an illustration. As mentioned
above, the empirically found distributions are approximately Gaussian.

At this point we can clearly formulate the main message of this
paper: In our model the presence of an initial peak of the friction
force, meaning that the maximum static friction is larger than the
dynamic friction, is the result of an initial {\em coherence} in the
strain distribution of the interlocked asperities.  The height of the
peak depends on the distribution of threshold lengths (or loop sizes,
see Fig.\ref{fig:hysteresis}).  After displacement of
the order of few times the average threshold length the initial
coherence is forgotten, the strains get out of phase, and as a
consequence the friction force decays to the dynamical friction
force. While we assume a complete initial coherence in our model, this
is not absolutely necessary: Decreasing the level of initial coherence
still results in a peak of the friction force, although with a
decreased height (results not shown here).


Now we show that the ratio of static and dynamical friction
coefficients does not change under a rescaling of loop sizes
$\ql \rightarrow a \ql$. With the probability distribution
\begin{equation}
  p(\ql) \rightarrow a p(a\ql),
\end{equation}
one gets according to eq.~(\ref{eq:fs}) that the elastic restoring
force at displacement $s$ and hence also its maximum value (static
friction force) and its average value (dynamical friction force)
transform as
\begin{equation}
f(s) \rightarrow \frac{1}{a} f(as).
\label{eq:scaling}
\end{equation}
Therefore the ratio between static and dynamic friction remains invariant.

As mentioned above, the
probability distribution $p(\ql)$ is approximately Gaussian in many
cases \cite{Greenwood1992}, {\em i.e.} 
completely characterized by its first and second moments, $\avrg{\ql}$
and $\avrg{\ql^2}$. As under a rescaling $\ql \rightarrow a \ql$ the
first and second moments scale differently, $\langle\ql\rangle
\rightarrow a \langle \ql\rangle$ and $\langle\ql^2\rangle
\rightarrow a^2 \langle \ql^2\rangle$, but the ratio $\muS/\muD$
remains invariant, it cannot depend on $\langle\ql\rangle$ and
$\langle\ql^2\rangle$ separately, but only on the invariant
combination $\langle\ql^2\rangle/
 \langle\ql\rangle^2$:
\begin{equation}
\frac{\muS}{\muD} =
g\left(\frac{\avrg{\ql^2}-\avrg{\ql}^2}{\avrg{\ql}^2}\right). 
\label{eq:ratio}
\end{equation}
The numerical analysis shows that $g$ is a decreasing function of its
argument (this tendency can be seen in fig.~\ref{fig:fs}a).  Therefore
the extreme case, where the width of $p(\ql)$ is zero
[$p(\ql)=\delta(\ql-\qlc)$] gives an upper bound on the ratio of the
friction coefficients.  This case is special, as coherence never gets
lost: All springs get stretched up to $\qlc$, break simultaneously and
are replaced by fresh, unstrained springs, which again get stretched
up to $\qlc$ and so forth. (The mathematical reason why our proof that
the force converges to the average value is not valid in this case is
that the $\delta$ function is not Riemann-integrable.)  Because of the
lack of decoherence, the model gives stick-slip motion. However, for
dry solid friction this is unphysical: Any small randomness for
example of the times, when new interlockings form, would have the
effect, that the springs ultimately get out of phase.  Then the
sawtooth-oscillations of the force $f(s)=A\, n_{\rm p}\, k \,
t(\qlc,s)$ would be damped similar to the oscillations shown in
fig.~\ref{fig:fs}a and would converge to the average value $\avrg{f} =
A\, n_{\rm p}\, k \, \qlc /2$, which we still identify with the
dynamic friction. Obviously it is half the maximal value of $f(s)$,
which we identify with the static friction. This gives an upper bound
of 2 for the ratio between the friction coefficients in
eq.~(\ref{eq:ratio}). Note that the upper bound would increase if time
strengthening were taken into account.
  
Together with the lower bound obtained if the relative width of
$p(\ql)$ tends towards infinity, we conclude that the model restricts
the ratio of the friction coefficients to the interval
\begin{equation}
1 \leq \frac{\muS}{\muD} \leq 2.
\end{equation}
Actually, for Gaussian distributions, a more stringent lower bound,
approximately 1.098, can be given. This value is obtained, when the
argument of $g$ in eq.~(\ref{eq:ratio}) tends to infinity, which
corresponds to the limit $\width/\ql_{\rm c} \rightarrow \infty$. Then
the force $f(s)$ approaches the one obtained for $\ql_{\rm c}=0$, {\em
i.e.}, for $p(\ql) \propto \exp[-\ql^2/2 (\width)^2]$. For this case the
ratio of the friction coefficients is approximately $1.098$ (see
fig.~\ref{fig:fs}b) independent of $\width$, due to
eq.~(\ref{eq:scaling}).  Lower values can be obtained if $p(\ql)$ has
a power law tail, \eg, $p(\ql) \propto 1 / [1 + (\ql / \qlc)^3]$, as
shown in fig.~\ref{fig:fs}a.

There is another interesting consequence of this theory. According to
the theory of Bowden and Tabor \cite{Bowden1958}, the microscopic contact
area of the pinning sites adjusts quickly by plastic flow such that the
local stress drops to the yield threshold $\yieldstress$ of the
material. Then the normal load is equal to the real contact area times
$\yieldstress$:
\begin{equation}
f_{\rm n} = A n_{\rm p} \yieldstress \alpha \avrg{\ql^2},
\end{equation}
where we assumed that a pinning site of loop size $\ql$ contributes
$\alpha \ql^2$ with a constant geometry factor $\alpha$ of order 1
to the real contact area. Combining this with eq.~(\ref{eq:average})
\begin{equation}
f_{\rm t} = \muD f_{\rm n} = \frac{k}{2} A n_{\rm p} \avrg{\ql}
\end{equation}
one finds that
\begin{equation}
\frac{\muS}{\muD} = g\left(\frac{A n_{\rm p}}{f_{\rm n}}
\frac{k^2}{4 \alpha \yieldstress \muD^2}  \right).
\end{equation}
As the friction coefficients should be independent of $f_{\rm n}$, 
we conclude that the number of pinning sites, $A \, n_{\rm p}$,
increases proportional to the normal load. This argument is not
entirely compelling, as the pinning strength $\ql$ needs not be
directly related to the microscopic contact area, and also the spring
constant $k$ might depend on the normal force.

In this work we presented a simple model of dry friction, which
explains why static friction force can be larger than
dynamic friction force, in terms of the concept of {\it coherence}.

\acknowledgments
The authors would like to thank L. Brendel and H. Hinrichsen for
fruitful discussions. This work was done within SFB 445
``Nano-particles from the Gasphase: Formation, Structure, Properties''.


\end{document}